\documentclass[apj]{emulateapj}   
\usepackage{graphicx}
\usepackage{multirow}

\usepackage{amsmath}
\usepackage{natbib}
\usepackage{hyperref}
\usepackage{physics}
\usepackage[usenames,dvipsnames]{color}  

\usepackage[normalem]{ulem}

\def\msun{{\rm ~M}_{\odot}}
\def\rsun{{\rm ~R}_{\odot}}
\def\zsun{{\rm ~Z}_{\odot}}
\def\gpy{{\rm ~Gpc}^{-3} {\rm ~yr}^{-1}}
\def\kms{{\rm ~km} {\rm ~s}^{-1}}

\definecolor{Orange-red}{rgb}{1.0, 0.27, 0.0}

\begin{document}

\title{The Origin of inequality: isolated formation of a $30+10\msun$ binary black-hole merger}

\author{
   A. Olejak\altaffilmark{1}, M. Fishbach\altaffilmark{2}, K. Belczynski\altaffilmark{1}, 
   D.E. Holz\altaffilmark{2}, J.-P. Lasota\altaffilmark{1,3}, M.~C. Miller\altaffilmark{4}, 
   T. Bulik\altaffilmark{5}
}

\affil{
   $^{1}$ Nicolaus Copernicus Astronomical Center, Polish Academy of Sciences,
          ul. Bartycka 18, 00-716 Warsaw, Poland\\ (aleksandra.olejak@wp.pl)\\
   $^{2}$ Enrico Fermi Institute, Department of Physics, Department of Astronomy \& 
          Astrophysics, KICP, University of Chicago, Chicago, IL 60637, USA\\
   $^{3}$ Institut d'Astrophysique de Paris, CNRS et Sorbonne Universit\'e,
          UMR 7095, 98bis Bd Arago, 75014 Paris, France\\
   $^{4}$ Department of Astronomy and Joint Space-Science Institute, University of 
          Maryland, College Park, MD 20742$-$2421, USA\\
   $^{5}$ Astronomical Observatory, Warsaw University, Al. Ujazdowskie 4, 00-478 
          Warsaw, Poland
}

\begin{abstract}

The LIGO/Virgo collaboration has reported the detection of GW190412, a black hole-black 
hole (BH-BH) merger with the most unequal masses to date\footnote{Another system with even 
more unequal mass components was  recently published by LIGO/Virgo: GW190814 
($m_1=23\msun$,$m_2=2.6\msun$), however it is not known whether it is a BH-BH or BH-NS 
merger \citep{2020LIGOMassGap}.}: $m_1=24.4$--$34.7\msun$ and 
$m_2=7.4$--$10.1\msun$, corresponding to a mass ratio of $q=0.21$--$0.41$ ($90\%$ 
probability range). Additionally, GW190412's effective spin was estimated to be 
$\chi_{\rm eff}=0.14$--$0.34$, with the spin of the primary BH in the range 
$a_{\rm spin}=0.17$--$0.59$. Based on this and prior detections, $\gtrsim 10\%$ of 
BH-BH mergers have $q\lesssim 0.4$. Major BH-BH formation channels (i.e., dynamics in 
dense stellar systems, classical isolated binary evolution, or chemically homogeneous 
evolution) tend to produce BH-BH mergers with comparable masses (typically with 
$q\gtrsim 0.5$). Here we test whether the classical isolated binary evolution channel 
can  produce mergers resembling GW190412.
We show that our standard binary evolution scenario, with the typical assumptions on input 
physics we have used in the past, produces such mergers. For this particular model of the 
input physics the overall BH-BH merger rate density in the local Universe ($z\sim0$) is: 
$73.5\gpy$, while for systems with $q<0.41$ the rate density is: $6.8\gpy$. The results 
from our standard model are consistent with the masses and spins of the black holes in 
GW190412, as well as with the LIGO/Virgo estimate of the fraction of unequal-mass BH-BH 
mergers.
As GW190412 shows some weak evidence for misaligned spins, we provide distribution of 
precession  parameter in our models and conclude that if among the new LIGO/Virgo detections 
the evidence of system precession is strong and more than $10\%$ of BH-BH mergers have large 
in-plane spin components ($\chi_{\rm p}>0.5$) then common envelope isolated binary BH-BH 
formation channel can be excluded as their origin. 

\end{abstract}


\section{Introduction}
\label{sec.intro}

The first confirmed double black hole (BH-BH) coalescence to be reported from the LIGO/Virgo 
O3 run, GW190412, differs from all previously announced BH-BH mergers in one important 
detail: it is the first BH-BH detection that has a mass ratio inconsistent with unity 
\citep{gw190412}. All ten BH-BH mergers announced by the LIGO/Virgo team from the O1 and O2 
observational campaigns were consistent with being equal-mass mergers~\citep{LIGO2019a,
LIGO2019b,2020ApJ...891L..27F}. In contrast, GW190412's component masses are 
$m_1=29.7^{+5.0}_{-5.3}\msun$ and $m_2=8.4^{+1.7}_{-1.0}\msun$, with a mass ratio of 
$q=0.28^{+0.13}_{-0.07}$ (median and 90\% symmetric credible interval) and a maximum mass 
ratio of $q=0.59$ ($99\%$ probability). The dimensionless spin of the primary (more massive) 
BH spin is estimated to be $a_{\rm spin1}=0.17$--$0.59$. The LIGO/Virgo collaboration also 
gave their constraints on the system effective spin parameter, which is expressed by the 
formula: 
\begin{equation} \label{eq:chi_eff}
\chi_{\rm eff}= \frac{m_1 a_{\rm spin1} \cos \theta_1+
m_2 a_{\rm spin2} \cos \theta_2}{m_1 + m_2}
\end{equation}
where $\theta_{i}$ is the angle between the individual BH spin $a_{\rm spini}$ and the 
system orbital angular momentum. The estimated value of the system effective spin parameter 
is $\chi_{\rm eff}=0.25^{+0.08}_{-0.11}$ ($90\%$ probability). 
The inferred BH-BH merger rate density from O1/O2 is $9.7$--$101\gpy$. From this and 
previous detections, $\gtrsim 10\%$ of BH-BH mergers have mass ratios $q<0.40$ 
\citep{gw190412}.

It is expected that merging BH-BH systems may form thorough several channels which are 
the classical isolated binary evolution channel~\citep{Bond1984b,Tutukov1993,Lipunov1997,
Voss2003,Belczynski2010a,Dominik2012,Kinugawa2014,Hartwig2016,Spera2016,Belczynski2016b,
Eldridge2016,Woosley2016,Stevenson2017,Kruckow2018,Hainich2018,Marchant2018,Spera2019,
Bavera2020}, the dense stellar system dynamical channel~\citep{PortegiesZwart2000,Miller2002a,
Miller2002b,PortegiesZwart2004,Gultekin2004,Gultekin2006,OLeary2007,Sadowski2008,
Downing2010,Antonini2012,Benacquista2013,Bae2014,Chatterjee2016,Mapelli2016,Hurley2016,
Rodriguez2016a,VanLandingham2016,Askar2017,ArcaSedda2017,Samsing2018,Morawski2018,
Banerjee2018,DiCarlo2019,Zevin2019,Rodriguez2018a,Perna2019,Kremer2020}, isolated multiple 
(triple, quadruple) systems  ~\citep{Antonini2017b,Silsbee2017,Arca-Sedda2018,LiuLai2018,
Fragione19}, mergers of binaries in galactic nuclei \citep{Antonini2012b,Hamers2018,
Hoang2018,Fragione2019b} and the chemically homogeneous evolution channel consisting of 
rapidly spinning stars in isolated binaries ~\citep{deMink2016,Mandel2016a,Marchant2016,
Buisson2020}. 

In those formation scenarios BH-BH systems typically form with comparable-mass 
components ($q\gtrsim0.5$). These predictions are challenged by GW190412.

In this study we demonstrate that in the isolated binary channel a small but significant 
fraction of systems lead to a BH-BH merger similar to GW190412. We provide a 
proof-of-principle example of an isolated binary that is both qualitatively and 
quantitatively indistinguishable from GW190412. We emphasize that we have implemented only 
one model, incorporating our best estimates of the physics and astrophysics which sets the 
evolution of stars in binary systems. We leave to future work a more extensive study, 
investigating a greater parameter space and exploring model uncertainties. Our results, 
when combined and contrasted with similar studies of other formation channels, suggest a 
plausible origin for GW190412.

\section{Calculations}
\label{sec.calc}

We use the population synthesis code {\tt StarTrack}~\citep{Belczynski2002,Belczynski2008a} 
to test the possibility of the formation of a BH-BH merger resembling GW190412. We employ 
the rapid core-collapse supernova (SN) engine NS/BH mass calculation~\citep{Fryer2012}, 
with weak mass loss from pulsational pair instability supernovae~\citep{Belczynski2016c}. 
We assume standard wind losses for massive stars: O/B star \cite{Vink2001} winds and LBV 
winds \citep[specific prescriptions for these winds are listed in Sec.~2.2 
of][]{Belczynski2010b}. 
BH natal spins are calculated under the assumption that angular momentum in massive  stars 
is transported by the Tayler-Spruit magnetic dynamo as adopted in the MESA stellar evolutionary 
code~\citep{Spruit2002}. Such BH natal spins are at the level of $a_{\rm spin} \sim 0.1$ (see 
\cite{Belczynski2020b}) and may be over-ridden if the immediate BH 
progenitor (WR) stars in close binaries (orbital periods $P_{\rm orb}<1.3$d) are subject 
to tidal interactions. In such cases we employ the scheme described in Sec.2.5 of 
\cite{Belczynski2020b}. For BH-WR, WR-BH and WR-WR binary systems with orbital periods in 
the range P$_{\rm orb}=0.1-1.3$d the BH natal spin magnitude is fit from WR star spun-up
 MESA models (see eq.15 of \cite{Belczynski2020b}), while for systems with P$_{\rm orb}<0.1$d 
the BH spin is equal to 1.0. BH spins may be increased by accretion in binary systems.   
We treat accretion onto a compact object during Roche lobe overflow 
(RLOF) and from stellar winds using the analytic approximations presented in 
\cite{King2001,Mondal2020}. We adopted limited $5\%$ Bondi accretion rate onto BHs 
during CE~\citep{Ricker&Taam, MacLeod2015a,MacLeod2017}. The estimate of Bondi accretion 
rate during CE phase is derived in Appendix B. 

The most updated description of {\tt StarTrack} is given in \cite{Belczynski2020b}. Here 
we use input physics from model M30 of that paper except for two important differences: 
First, instead of using the initial mass ratio distribution from \cite{Sana2012}, which 
allows only $q=0.1$--$1.0$, we now extend this distribution to lower mass ratios 
$q=q_{\rm min}$--$1.0$, where $q_{\rm min}$ is chosen in such a way that a star mass is 
allowed to reach the hydrogen burning limit $M_{\rm ZAMS}=0.08\msun$. Second, for cases 
in which we do not know whether we should apply thermal timescale RLOF or CE for systems 
with NS/BH accretors we use a specific diagnostic diagram to decide between thermal RLOF 
and CE (see Sec.~5.2 of \cite{Belczynski2008a}). In this single step of binary evolution 
we previously applied our older numerical approximation of the calculation of accretion 
onto NS/BH presented in~\cite{Belczynski2008b} instead of our newly adopted analytic 
approach~\citep{King2001,Mondal2020}. These two changes increase the estimated total BH-BH 
merger rate in the local Universe ($z\sim0$) from $43.7\gpy$ (model M30.B; 
\cite{Belczynski2020b}) to $73.5\gpy$ (this study; see below).

\section{Example of Evolution}
\label{sec.evol}

\begin{figure}
\hspace*{-0.4cm}
\includegraphics[width=0.5\textwidth]{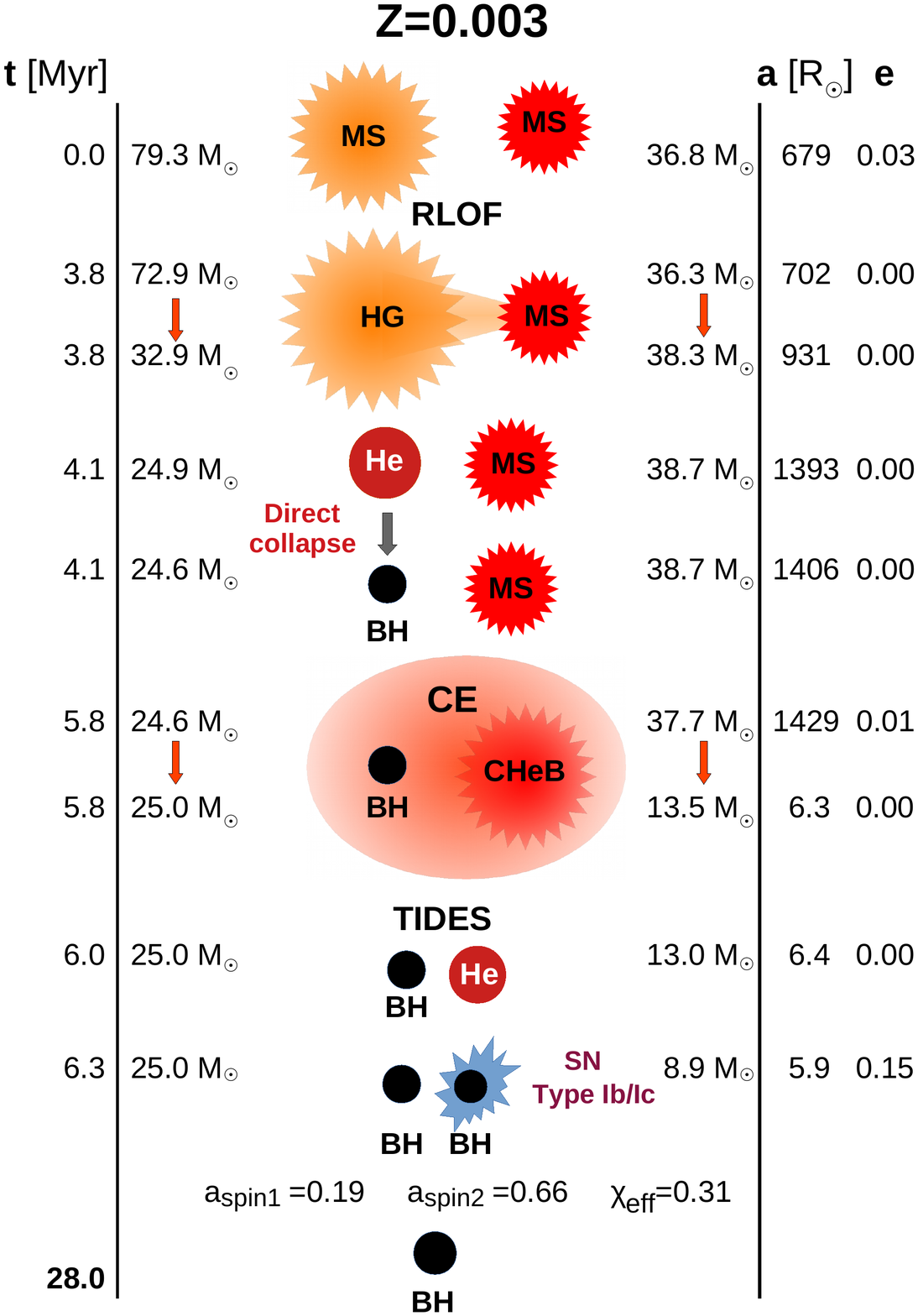}
\caption{
Evolution of an isolated binary system that produces a BH-BH merger resembling GW190412 
(see Sec.~\ref{sec.evol} for details). MS: main sequence star, HG: Hertzsprung gap star, 
CHeB: core helium burning star, He: naked helium star, BH: black hole, RLOF: Roche lobe 
overflow, CE: common envelope.
}
\label{fig.evol1}
\end{figure}

In Figure~\ref{fig.evol1} we present an example of the evolution of a binary system which 
leads to the formation of close BH-BH system consistent with the parameters estimated for 
GW190412. This system was picked from the most populated formation channel of BH-BH 
mergers with $q<0.41$ (see Tab.~\ref{tab.rates}). This system has both BH masses and 
primary BH spin $a_{\rm spin1}$ within the range of $90\%$ uncertainties given by 
\cite{gw190412}. 

This system, with initial primary mass $\sim 79\msun$ and secondary mass 
$\sim 37\msun$, is formed in a low metallicity environment $Z=0.003$ ($\sim 0.1\zsun$) 
with an initial separation of $a\sim 680\rsun$ and eccentricity $e\sim 0.03$. When the 
more massive star leaves the main sequence, the system circularizes ($e=0.0$) 
at the onset of the stable RLOF phase, during which the donor (primary star) loses a 
significant amount (over $50\%$) of its mass. After finishing its nuclear evolution, the 
primary undergoes direct collapse and forms a first BH with no natal kick and no 
associated supernova explosion. After the secondary leaves the main sequence and becomes 
a core helium burning giant, the system enters a CE phase during which the secondary loses 
its H-rich envelope. The system separation decreases to only $a\sim 6\rsun$. After CE, the 
secondary is a massive naked helium WR star. The binary separation is so small 
that the secondary is subject to strong tidal interactions and is spun up. At time $t=6.3$ Myr 
since the start of the evolution, the secondary explodes as a Type Ib/c supernova (mass 
ejection of $\sim 3.0\msun$; 3D natal kick of $v_{\rm kick}=98\kms$) and forms a second 
BH. Due to the small orbital separation, the two BHs, now with a mass ratio of 
$q=0.36$, merge in just $\sim 21.7$ Myr.

The first BH forms with a spin $a_{\rm spin1}=0.13$ (calculated from MESA single stellar 
models with \cite{Spruit1999} angular momentum transport; see Fig.~2 of \cite{Belczynski2020b}) 
that is perfectly aligned with the binary angular momentum ($\theta_1=0\deg$). Had we adopted 
more efficient angular momentum transport in stars \citep{Fuller2019a,Fuller2019b,Ma2019} than 
employed in the standard MESA 
then primary BH spin would change to $a_{\rm spin1}\sim 0.01$. This BH accretes in CE and 
during stable RLOF from its companion ($\sim 0.4\msun$) and increases its spin to 
$a_{\rm spin1}=0.19$. The second, lower mass, BH forms with spin $a_{\rm spin2}=0.66$ that 
is slightly misaligned by its natal kick to $\theta_2=5\deg$. The spin magnitude is 
obtained from rapidly spinning MESA naked helium star models with spins that correspond to
a tidally locked star for a given orbital period in our binary models (see Eq.~15 of 
\cite{Belczynski2020b}). The effective spin parameter of this BH-BH merger is 
$\chi_{\rm eff}=0.31$, within the LIGO/Virgo range for GW190412 ($0.14$--$0.34$). It is 
noted that for the virtually aligned geometry of BH spins with binary angular momentum in 
this example we do not expect any precession. Yet, there seems to be marginal evidence for 
precession in GW190412. We provide a discussion of precession in Section~\ref{sec:concl} 
and Appendix A.

One might be tempted to identify phase 4 (just before CE in Fig.~\ref{fig.evol1}) of the 
evolution of our binary system with high-mass X-ray binaries of the Cyg X-1 type 
($M_{\rm BH}=14.8\msun$, O star companion  $M_{\rm O}=19.2\msun$ and orbital period of 
$P_{\rm orb}=5.6$d\footnote{\url{https://universeathome.pl/universe/blackholes.php}}; this
corresponds to a semi-major axis of $a=43\rsun$). However, Cyg X-1 is an active system (it 
accretes from a wind), which implies an orbital separation that is too tight to allow 
survival of the subsequent CE phase~\citep{Belczynski2012b}.  If it instead undergoes a 
stable RLOF~\citep{Pavlovskii2015,Pavlovskii2017} then the orbit will widen beyond the 
limit ($a \sim 50\rsun$) for two BHs to merge within a Hubble time. We note that BH-BH 
progenitors in our simulations are initially very wide ($a \gtrsim 1000\rsun$) binaries 
so they can successfully survive the CE phase~\citep{deMink2015}.

\begin{figure}
\hspace*{-0.4cm}
\includegraphics[width=0.5\textwidth]{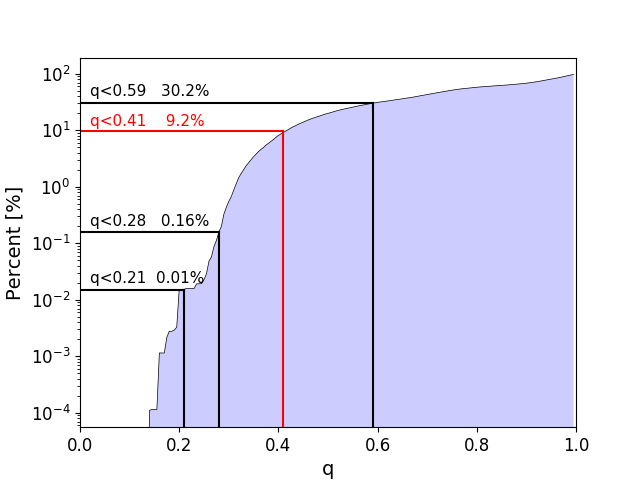}
\caption{
Cumulative fraction of merging BH-BH systems with mass ratio smaller than $q$ in the local 
Universe ($z\sim0$). Fractions for selected mass ratios $q<0.21$ ($0.01\%$), $q<0.28$ 
($0.16\%$), and $q<0.59$ ($30.2\%$) are marked with black lines. The red line marks $q<0.41$, 
indicating that $9.2\%$ of our simulated binary mergers at $z\sim0$ are consistent with 
the $90\%$ upper limit on $q$ for GW190412. 
}
\label{fig.q}
\end{figure}

\section{Population of low-$\lowercase{q}$ BH-BH mergers}
\label{sec.q}

Our simulation results in a $z \sim 0$ population of merging BH-BH systems with a local 
rate density of ${\cal R}_0=73.5\gpy$. The cumulative distribution of mass ratios for 
these mergers is presented in Figure~\ref{fig.q}. In this model the majority of BH-BH 
mergers ($\sim 80\%$) have large mass ratios ($q>0.5$), consistent with previous 
results~\citep{Belczynski2016b}. Here we focus on the tail of the distribution extending 
to more extreme mass ratios. Our model predicts very few systems with mass ratios smaller 
than the average value reported for GW190412: $0.16\%$ of binaries have $q<0.28$. However, 
we report a more significant fraction of systems with mass ratios smaller than the $90\%$
upper bound on GW190412: $9.2\%$ at $q<0.41$. This fraction becomes significantly higher 
for the $99\%$ upper bound on GW190412: $30.2\%$ at $q<0.59$.

In Table~\ref{tab.rates} we show evolutionary sequences that lead to the formation of 
BH-BH mergers with small mass ratios: $q<0.41$. We list the merger rate density arising 
for typical evolutionary sequences. These are $z\sim 0$ rate densities and are 
subpopulations of the overall local BH-BH merger population (${\cal R}_0=73.5\gpy$). 
The table presents merger rate densities for BH-BH systems which are increasingly 
constrained to resemble GW190412: 
\begin{enumerate}

\item ${\cal R}_1$: $q<0.41$,

\item ${\cal R}_2$: $q<0.41$ and $24.4<m_1/\msun<34.7$ and $7.4<m_2/\msun<10.1$,

\item ${\cal R}_3$: $q<0.41$ and $24.4<m_1/\msun<34.7$ and $7.4<m_2/\msun<10.1$
      and $0.17<a_{\rm spin,a}<0.59$,

\item ${\cal R}_4$: $q<0.41$ and $24.4<m_1/\msun<34.7$ and $7.4<m_2/\msun<10.1$
      and $0.17<a_{\rm spin,a}<0.59$ and $0.14<\chi_{\rm eff}<0.34$.

\end{enumerate}
The overall rate of systems with $q<0.41$ is ${\cal R}_1=6.8\gpy$, which corresponds to 
$\sim 10\%$ of our overall predicted local merger rate density of BH-BH systems 
(${\cal R}_0=73.5\gpy$). This is consistent with the LIGO/Virgo estimate of the 
fraction of low mass ratio systems as inferred from the detection of GW190412 combined 
with previous detections. We emphasize that Figure~\ref{fig.q} shows the distribution of
the mass ratios for {\em all}\/ merging binaries in the local Universe, which may be 
different from the distribution of {\em detected}\/ binaries since it does not 
incorporate gravitational-wave (GW) selection effects. This is not expected to lead 
to a significant effect in the case of mass ratio distributions~\citep[e.g., see Fig.~4 
of][]{2020ApJ...891L..27F}. In addition, it is to be noted that the LIGO/Virgo estimate 
of $\gtrsim 10\%$ of binaries having $q\lesssim 0.4$~\citep{gw190412} is for the true 
(intrinsic) population, not the detected population. This estimate is thus directly 
comparable to the results from Fig.~\ref{fig.q}.

To produce a low mass ratio system with a primary BH as massive as $30\msun$, a progenitor
binary needs to have {\em (i)} one very massive component ($M_{\rm ZAMS} \gtrsim 70\msun$), 
and {\em (ii)} rather low initial stellar mass ratio ($q_{\rm ZAMS}<0.5$). In addition, the 
progenitor binary needs to have low metallicity $Z\lesssim 10\%\zsun$ ($\lesssim 0.002$)
~\citep{Belczynski2010b}. These systems are uncommon, leading to a dearth of small mass-ratio 
BH-BH mergers such as GW190412.

\begin{table}
\caption{Evolutionary channels for $q<0.41$ BH-BH mergers}
\begin{tabular}{cccccc}
\hline\hline
No. & Evolutionary history$^{a}$ & ${\cal R}_1^{b}$ & ${\cal R}_2^{c}$ & ${\cal R}_3^{d}$ & ${\cal R}_4^{e}$ \\
\hline\hline
1 & RLOF1 BH1 CE2 BH2       & 5.90 & 0.49 & 0.11 & 0.11 \\
2 & RLOF1 BH1 RLOF2 CE2 BH2 & 0.76 & 0.04 & 0.01 & 0.01 \\
3 & RLOF1 BH1 CE2 RLOF2 BH2 & 0.02 & 0.01 & 0.00 & 0.00 \\
4 & OTHER CHANNELS          & 0.11 & 0.00 & 0.01 & 0.00 \\
\hline
  & All                     & \bf{6.79} & 0.54 & 0.13 & 0.11 \\
\hline
\hline
\end{tabular}
$^{a}$:
RLOF: stable Roche lobe overflow, CE: common envelope, BH: black hole
formation, 1: indicates primary (initially more massive star), 2:
secondary star being donor in RLOF or CE.\\
$^{b}$: Merger rate density ($\gpy$) for systems with $q<0.41$.\\
$^{c}$: above and $24.4<m_1/\msun<34.7$ and $7.4<m_2/\msun<10.1$.\\
$^{d}$: above and $0.17<a_{\rm spin,a}<0.59$.\\
$^{e}$: above and $0.14<\chi_{\rm eff}<0.34$.\\
\label{tab.rates}
\end{table}

\section{Discussion and Conclusions}
\label{sec:concl}

The existence of unequal mass binary black holes is to be expected within the isolated 
binary evolution formation scenario. The mass ratios of such systems were initially 
investigated by \cite{Bulik2004}. They found that in the standard scenario one expects 
BH-BH systems with high mass ratios above $0.7$ to dominate; however, varying the 
efficiency of the common envelope evolution phase leads to the formation of systems with  
mass ratios less than $0.5$. Although our knowledge of binary evolution and BH-BH 
formation has subsequently improved, this result appears robust and remains valid. 
\cite{Dominik2012} have shown the distribution of mass ratios of BH-BH systems in their 
Figure~9. They find that for sub-solar metallicity a significant fraction of these mergers 
have mass ratio less than $0.5$. An additional hint for the existence of unequal mass 
BH-BH systems from isolated binary evolution comes from the analysis of the future 
evolution of Cyg X-3 \citep{Belczynski2013}. This system will lead to formation of either 
a BH-NS or BH-BH binary; in the latter case, the mass ratio is expected to be below $0.6$.  
Systems with BH masses similar to GW190412 are also found in results from isolated 
binary evolution calculations by other groups (e.g., see Fig.5 of \cite{Eldridge2016}).

The formation channel of GW190412 was considered by~\cite{DiCarlo2020}, both through 
dynamical formation in open clusters and through the classical isolated binary evolution 
channel as discussed here. That group finds that systems like GW190412: ``can be matched 
only by dynamical BH-BH born from metal-poor progenitors, because isolated binaries can 
hardly account for its mass ratio in our models.''
Unlike them, we find that systems like GW190412 are naturally formed by isolated binaries 
in a small but significant fraction of systems. Note also that the model that we use to 
account for the formation of GW190412 has also been used to explain the merger rates, 
masses, and low effective spins of the full O1/O2 LIGO/Virgo BH-BH merger sample
~\citep{Belczynski2020b}. 

\cite{mandel2020} have questioned the LIGO/Virgo conclusion that the non-negligible 
positive effective spin parameter for GW190412 has its origin from a moderate/high spin of 
the primary (more massive) BH in GW190412 ($a_{\rm  spin1}=0.17-0.59$). Instead, 
\cite{mandel2020} point out that in the classical isolated binary evolution scenario some 
second-born BHs may form from tidally spun up helium stars, and that the resulting BHs are 
expected to have high spins. Using priors consistent with this, they perform an 
alternate analysis of GW190412 which finds that the primary BH has negligible spin 
($a_{\rm spin1} \sim 0$) while the secondary BH has high spin ($a_{\rm spin2}=0.64-0.99$). 
This possibility is also consistent with our results: we find that in $\sim 30\%$ of local 
BH-BH mergers with $q<0.41$, tidal interactions are strong enough to produce a lower-mass 
BH with spin $a_{\rm spin2}>0.64$. For example, in Figure~\ref{fig.evol1} we show a system 
that forms a very close ($a\sim 4\rsun$) binary with a BH and a naked helium star (this is 
the evolutionary phase just prior BH-BH formation). This naked helium star is subject to 
tidal spin-up, and instead of forming a slowly spinning BH, it forms a rapidly spinning BH 
($a_{\rm spin2}=0.66$). However, in contrast with \cite{mandel2020} we do not assume that 
the primary BH spin is negligible. Instead, we calculate the natal BH spins (if not affected 
by tides) from single stellar models allow for spin increase due to accretion during binary 
mass transfer phases (see Sect. \ref{sec.calc}). The primary BH spins are found to 
be small, but not negligible. For the case shown in Figure~\ref{fig.evol1}, the natal 
primary BH spin is $a_{\rm spin1}=0.11$ and then  it is increased to $a_{\rm spin1}=0.19$ 
through accretion in common envelope event. Since both spins are closely aligned with 
the binary angular momentum (the secondary is slightly misaligned due to a small natal kick, 
to $\theta_2=5\deg$), the effective spin parameter of this system is $\chi_{\rm eff}=0.31$, 
which is consistent with the upper end of the LIGO/Virgo $90\%$ probability estimate for 
GW190412.

GW190412 shows some weak evidence for misaligned spins, with a non-zero precession parameter: 
$\chi_{\rm p}=0.15-0.49$ ($90\%$ credible limits). In this system, the amount of observed 
precession is consistent with noise (see Fig. 6 of \citealt{gw190412}), and the mild 
preference for $\chi_p>0$ disappears when the gravitational-wave data is reanalyzed with 
different priors on the spin magnitudes \citep{2020arXiv200611293Z}. Nevertheless, it is 
interesting to explore whether a clear observation of precession would be consistent with 
our models. In our evolutionary example (see Fig.~\ref{fig.evol1}) we do not expect to 
produce any precessing systems as both BHs are almost fully aligned with the binary angular 
momentum. Some degree of misalignment would appear in our model if, for example, we added a 
larger natal kick at the formation of the second BH. At this point the binary is so tight 
that even a large kick would have only a small chance to disrupt this binary. The small 
natal kick applied to the second BH formed through partial fallback results from the simple 
assumption that natal kicks scale inversely with the amount of the fallback~\citep{Fryer2012} 
but little is known little is known about BH natal kicks~\citep{Repetto2015,Mandel2016b,
Belczynski2016a,Repetto2017,Gandhi2020}.
We have estimated the precession parameter for all the BH-BH mergers produced by our model 
(see Appendix). The cumulative distribution of $\chi_{\rm p}$ (Fig.~\ref{fig.Xp_cum}) shows 
that BH-BH mergers are dominated by low precession parameters values for our standard model 
(small natal BH kicks). We calculated several additional models adopting high BH natal kicks, 
and different approach to tidal spin up of BH progenitors to be able to provide an exclusion 
statements. 
If an analysis of the LIGO/Virgo BH-BH population finds that more than $10\%$ of BH-BH 
mergers have large in-plane spin components ($\chi_{\rm p}>0.5$) then common envelope 
isolated binary BH-BH formation channel can be excluded as their origin. This conclusion is 
valid if {\em (i)} stars in binaries are born with aligned spins and {\em (ii)} angular 
momentum transport in massive stars is efficient (driven by magnetic dynamo) producing low 
natal BH spins ($a_{\rm spin}<0.2$), unless BH progenitor stars are subject to tidal spin-up. 
Furthermore, this conclusion is independent of the black hole natal kick model or the action 
of tides on Wolf-Rayet stars in close binaries. 
A similar statement can be made for possible future signals from highly mass asymmetric 
BH-BH systems with large $\chi_{\rm{eff}} \gtrsim 0.5$. We show distributions of effective 
spin parameter for the overall local BH-BH mergers and the low mass ratio BH-BH sub-population 
(Fig. \ref{fig.Xeff}). This figure indicates that the effective spin values for low-q 
sub-population are systematically smaller and limited to $\abs{\chi_{\rm{eff}}} \leq 0.5$.

We have shown that the isolated classical binary evolution channel can form binaries 
similar to GW190412. This is an important explicit proof-of-principle demonstration that 
the event GW190412 may be the result of isolated evolution. Furthermore, Fig.~\ref{fig.q} 
shows that the detection of a binary with a mass ratio of $q\lesssim 0.4$ is to be 
expected within the current GW sample, since this sub-population constitutes $\sim 10\%$ 
of the total population. We find that, if GW190412 formed via the classical isolated 
binary channel, it likely evolved from a low-metallicity ($Z<10\%\zsun$) progenitor system 
with initial mass ratio $q<0.5$ between the two massive stars, but that otherwise the system 
followed an evolutionary path that is typical of the majority of BH-BH mergers
~\cite{Belczynski2016b}. Over the coming years the population of GW BH-BH mergers is 
expected to grow to many hundreds of detections. These will facilitate detailed population 
studies, including a determination of the distribution of mass ratios. While the existing 
population of BH-BH mergers can be explained using classical isolated binary evolution, the 
discovery of a large population of binaries with mass ratio $q<0.2$ would pose a significant 
challenge to our models.

\acknowledgements
We would like to thank anonymous referee for their useful comments.
KB and AO acknowledge support from the Polish National Science Center (NCN) grant
Maestro (2018/30/A/ST9/00050). JPL was supported in part by the French Space Agency
CNES. TB was supported by TEAM/2016-3/19 grant from FNP.
DEH was supported by NSF grant PHY-1708081, as well as the Kavli Institute for Cosmological 
Physics at the University of Chicago through an endowment from the Kavli Foundation.
DEH also gratefully acknowledges support from the Marion and Stuart Rice Award.
MCM thanks the Radboud Excellence Initiative for supporting his stay at Radboud University.

\bibliography{biblio}

\clearpage
\section{Appendix A: Precession Parameter}
\label{sec:appendix}

The LIGO/Virgo Collaboration gave an estimate of the GW190412 precession parameter $\chi_{\rm p}$ 
which is spin-dependent parameter expressed by the formula  \citep{Schmidt_2015,Gerosa2020}:
\begin{equation} 
\label{eq: chi_p}
\chi_{\rm p} = \mbox{max} \left [a_{\mbox{spin1}} \sin \theta_1, a_{\mbox{spin2}} \sin \theta_2 \frac{q(4q + 3)}{(4 + 3q)} \right ]  
\end{equation} 
The value of $\chi_{\mbox{p}}$ given by LIGO/Virgo is in the range of $0.15-0.49$ ($90\%$ 
credible limits). This is unique among the other BH-BH merger detections for which the 
precession parameter was uninformative, and consistent with $\chi_{\rm p} = 0$ (corresponding 
to perfectly aligned spins). GW190412 shows weak evidence for precession, we note that the 
measurement remains inconclusive, and small values of $\chi_{\rm p} < 0.1$ cannot be ruled out
~\citep{gw190412,2020arXiv200611293Z}. Although $\chi_{\rm p}$ is poorly measured for individual 
gravitational-wave events, combining multiple observations will reveal the population distribution 
of $\chi_{\rm p}$. This will provide a powerful test of our models, as we discuss below.

We calculated distributions of precession parameter values for our standard model and its several 
variations. 

\begin{figure*}
\hspace*{-0.4cm}
\includegraphics[width=1.0\textwidth]{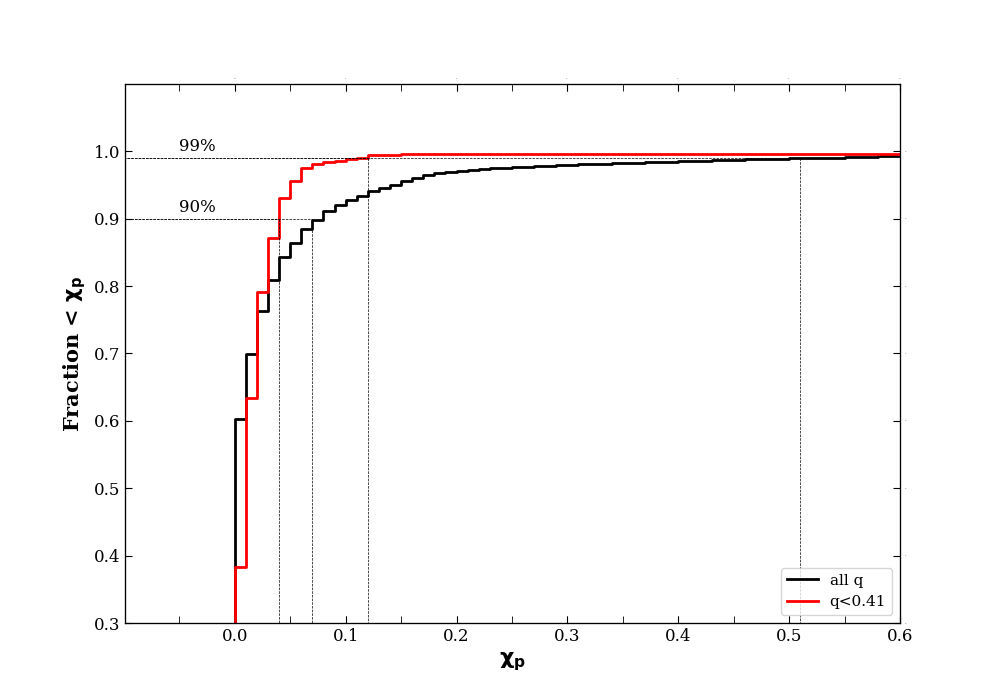}
\caption{
Cumulative distribution of precession parameter $\chi_{\mbox{p}}$ of BH-BH mergers in 
the local Universe ($z\sim0$). Black line - overall BH-BH population; red line - 
sub-population of BH-BH mergers with mass ratio $q<0.41$. Results for standard 
model: Spruit-Tayler BH spins + natal kicks lowered by fallback and partial
tidal interactions.
}
\label{fig.Xp_cum}
\end{figure*}

The cumulative distribution of $\chi_{\rm p}$ for our standard model is shown in Figure 
~\ref{fig.Xp_cum}. For the overall BH-BH population merging at $z \sim 0$ as well as for the low 
mass ratio sub-population ($q<0.41$), the distribution is dominated by low precession parameters 
values: $90\%$ of overall BH-BH binaries have precession parameter $\chi_{\rm p}<0.07$ 
while $99\%$ have $\chi_{\rm p}<0.51$. Low mass ratio BH-BH mergers have even lower values: 
$90\%$ of the systems have $\chi_{\rm p}<0.04$ and $99\%$ have $\chi_{\rm p}<0.11$. The 
reason of the difference between those two populations is the relation between natal kicks 
and the mass of resulting compact object. Less massive BHs usually get higher natal kicks, so 
precession is more likely in mergers with low mass BHs. In the low mass ratio mergers, 
one of the black holes is always massive, and is formed through direct collapse (without a
SN explosion). In contrast, in the overall BH-BH population, there are cases of mergers with 
two low-mass BHs that may form with high natal kicks. High natal kicks increase the degree 
of misalignment and subsequently increase $\chi_{\rm p}$.

\begin{figure*}
\hspace*{-0.4cm}
\includegraphics[width=1.0\textwidth]{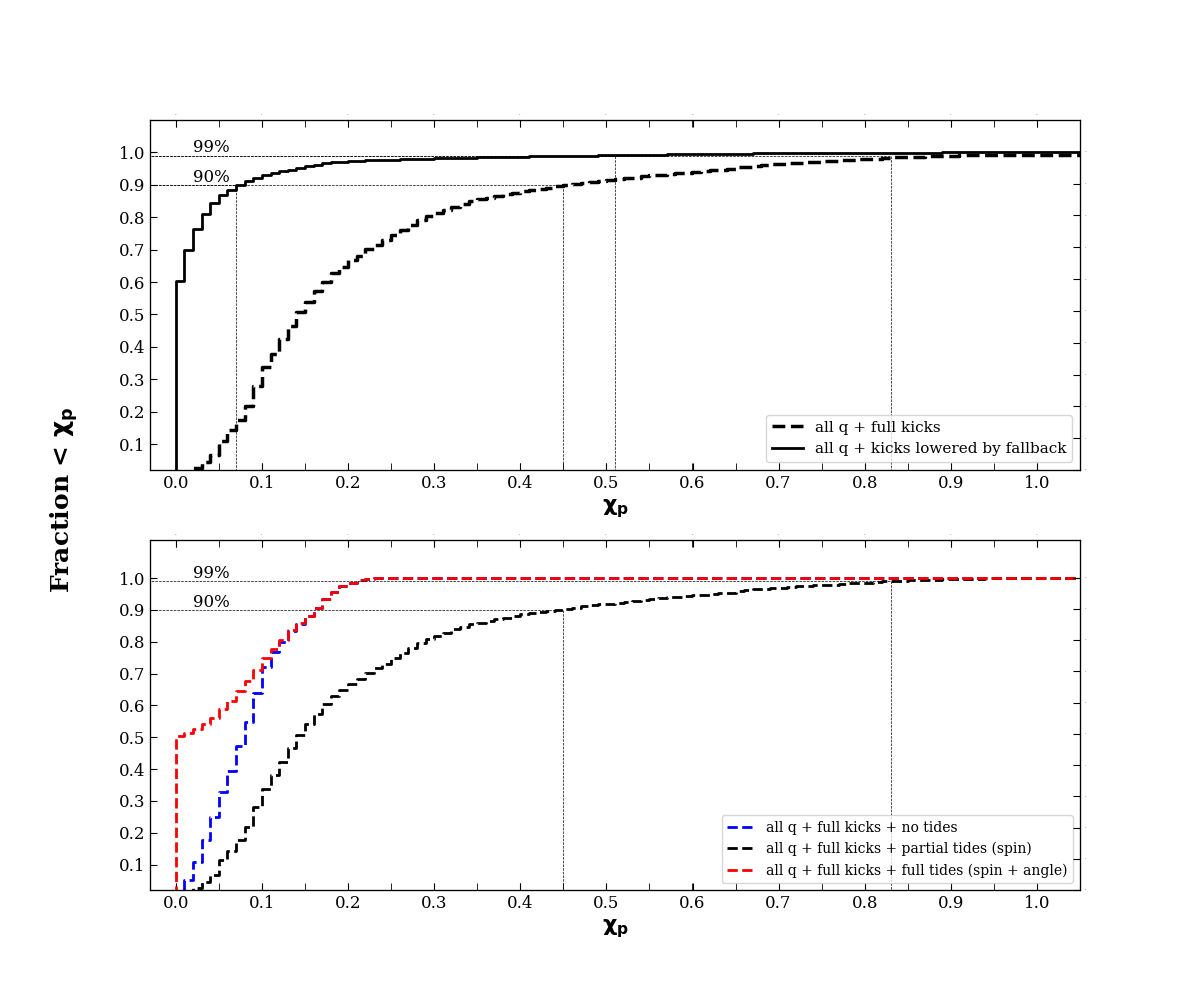}
\caption{
Cumulative distribution of precession parameter $\chi_{\mbox{p}}$ of BH-BH mergers in 
the local Universe ($z\sim0$). \textbf{Top:} solid black line - overall BH-BH 
population with standard natal kicks lowered by fallback; dashed black line - overall 
BH-BH population with full natal kicks; \textbf{Bottom:} blue dashed line - overall 
BH-BH population with full natal kicks and no tidal interactions on Wolf-Rayet stars; 
black dashed line - overall BH-BH population with full natal kicks and partial tidal 
interactions on Wolf-Rayet stars (spin magnitude); red dashed line - overall BH-BH 
population with full natal kicks and full tidal interactions on Wolf-Rayet stars (spin 
magnitude and angles);
}
\label{fig.Xp_cum_nf}
\end{figure*}

To test the maximum allowed level of precession in our isolated binary evolution model, we 
increase BH natal kicks to the high speeds observed for single pulsars in the Galaxy 
(Maxwellian distribution with $\sigma=265\kms$; \cite{Hobbs2005}), and apply these natal 
kicks to all BHs independent of their mass. In this model, the distribution of precession 
parameters shifted to higher values, with $90\%$ of all BH-BH mergers (any $q$) having 
$\chi_{\rm p}<0.43$ and $99\%$ having $\chi_{\rm p}<0.82$ (see top panel of 
Fig.~\ref{fig.Xp_cum_nf}). Note that our standard model employs BH natal kicks decreased by 
fallback, and in practice, massive BHs ($M \gtrsim 10-15\msun$) do not receive natal kicks. 
 
We have also tested the effect of tidal interactions between Wolf-Rayet star (an immediate 
BH progenitor in our models) and its massive companion on precession parameter. Three 
variants of approach to tides are shown in the bottom panel of Figure~\ref{fig.Xp_cum_nf}. 
Note that tides may change misalignment angles and BH natal spin magnitude affecting the 
value of $\chi_{\rm p}$. We perform this analysis on high natal BH kick model to maximize 
the effect of tides. We tested a variant with no tidal interactions on WR stars (no tides), 
a variant in which tides only affect spin magnitude (partial tides: our standard model 
approach), and a variant in which tides affect spin magnitude and cause alignment of
Wolf-Rayet star with binary angular momentum  (full tides; note that this star spin may be 
misaligned if earlier natal kick on the other star shifted binary angular momentum vector). 

We find that among these three drastically different approaches to tides, our standard model 
(partial tides) may be considered as an upper limit on $\chi_{\rm p}$ parameter value. For 
both no tides and full tides variants $90\%$ of systems have $\chi_{\rm p} \lesssim 0.2$ 
while $99\%$ of systems have $\chi_{\rm p} \lesssim 0.15$, which is much less than for the 
variant with partial tides (for $90\%$ $\chi_{\rm p}<0.43$ and for $99\%$ $\chi_{\rm p}<0.82$). 
The lower limit on $\chi_{\rm p}$ values in no tides variant is simply related to the fact 
that the BH spin magnitudes are not increased due to tidal interactions in the WR phase. In 
the case of full tides the sharp increase in fraction of systems near $\chi_{\rm p} \sim 0$ 
is generated due to the assumption about the WR star spin alignment with the system angular 
momentum so the part of the formula corresponded to a given BH (Eq.~\ref{eq: chi_p}) takes 
the value of zero. Those differences cause the removal of systems with high $\chi_{\rm p}$ 
from the distribution for no tides and full tides variants as contrasted with our partial 
tide model.

\begin{figure*}
\hspace*{-0.4cm}
\includegraphics[width=1.0\textwidth]{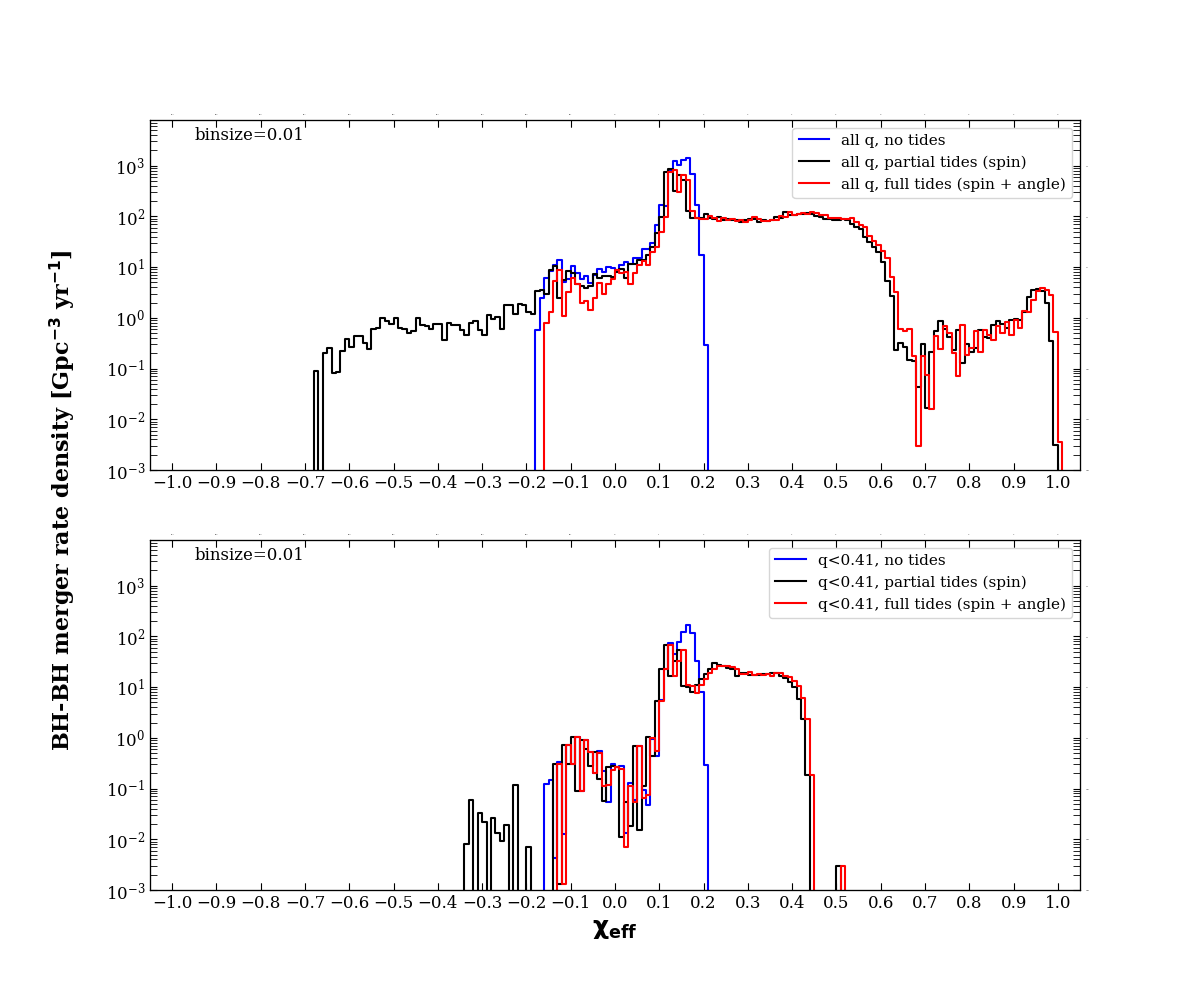}
\caption{
Distribution of effective spin parameter $\chi_{\mbox{eff}}$ of BH-BH mergers in 
the local Universe ($z\sim0$) for different approaches to tides: blue line - no tides; 
black line - partial tides; red line - full tides. \textbf{Top:} overall BH-BH population; 
\textbf{Bottom:} - sub-population of BH-BH mergers with mass ratio $q<0.41$. Results for 
standard model: Spruit-Tayler BH spins + natal kicks lowered by fallback.
}
\label{fig.Xeff}
\end{figure*}

In Figure \ref{fig.Xeff} we present effective spin $\chi_{\rm{eff}}$ distribution in our 
standard model for three different approaches to tides. In the top panel there is a distribution 
for overall BH-BH population merging at z$~\sim 0$ and in the bottom panel we show distribution 
for low mass ratio sub-population with $q<0.41$. Adopted tides approaches gives different results, 
especially for overall BH-BH population. In no tides approach both distributions (overall and low 
q) are similarly dominated by low effective spin parameter and the absolute value is limited to 
$\abs{\chi_{\rm{eff}}}<0.2$. In partial and full tides approaches the possible absolute value of 
effective spin widens to around $\abs{\chi_{\rm{eff}}}<0.5$ for low mass ratio sub-population 
while in overall population effective spin may take values up 1.0. This is caused by the fact that 
for the overall BH-BH population there are more possible evolutionary scenarios in which both 
objects could be the subject of WR-tides.

Based on our results we may conclude that if an analysis of the LIGO/Virgo BH-BH population 
reveals that more than $10\%$ of systems have high precession ($\chi_{\rm p} \leq 0.5$) then 
common envelope isolated binary BH-BH formation channel can be excluded as their origin. 
This conclusion is valid if {\em (i)} stars in binaries are born with aligned spins and 
{\em (ii)} natal BH spins are low ($a_{\rm spin}<0.2$) unless their progenitor stars are 
subject to strong tidal interactions, and is independent of the black hole natal kick model 
or the action of tides on Wolf-Rayet stars in close binaries. We note that we have assumed 
in all our simulations that stellar spins are aligned with the binary angular momentum at 
ZAMS, that only natal kicks at BH formation may misalign stellar/BH spins, and that only 
tidal interactions can realign stellar spins. A similar statement can be made for possible 
future signals from highly mass asymmetric BH-BH systems with large $\chi_{\rm{eff}} \gtrsim 0.5$. 
Distributions indicates that the effective spin values for low-q sub-population are systematically 
smaller and limited to $\abs{\chi_{\rm{eff}}} \leq 0.5$.

\section{Appendix B: Accretion during CE phase}
\label{sec:appendix_2}

Here we describe the procedure of calculating the accretion rate onto the BH during the 
CE phases. The procedure is based on equations 5.3-5.7 of \cite{Bethe1998} and equations 
A1-A10 from \cite{Belczynski2002}. 

In our calculations CE begins once BH companion (CE donor) expands beyond its Roche lobe 
and mass transfer is determinded to proceed on a dynamical timescale~\citep{Belczynski2008a}. 
CE evolution and accretion onto the BH ends when the donor’s envelope is ejected and the 
donor mass is reduced to the mass of its core. We use the following symbols: $\rm M_{\rm A}$ 
- mass of the BH, $\rm M_{\rm B}$ - mass of the donor, $\rm M_{\rm{B,core}}$ - mass of the 
donor's core, $A$ - orbital separation (semi-major axis).

First, we compare energy loss rate related to the accretion onto the BH and the rate of the 
orbital energy dissipation due to the dynamical friction of BH in the donor's envelope: 
\begin{equation}
\dot{E}_{\rm{acc}} =- \dot{E}_{\rm{orb}}
\label{eq.100}
\end{equation}
The formula for $\dot{E}_{\rm{acc}}$ is introduced by equations 5.3-5.7 in \citep{Bethe1998} 
and A1 \citep{Belczynski2002} while $\dot{E}_{\rm{orb}}$ is expressed by equation A2 
\citep{Belczynski2002}. Note, that $\dot{E}_{\rm{acc}}$ include mass accretion rate 
$\dot{M}_{\rm{A}}$ given by the Bondi-Hoyle-Lyttleton theory. Comparing the time derivatives 
of both energies we obtain first time independent differential equation which contains 
$\frac{dM_A}{dM_B}$ and $\frac{dA}{dM_B}$ (Eq. A3 of \cite{Belczynski2002}). 

Second, we compare the donor's envelope binding energy with the orbital energy, since 
CE is ejected on the expense of the binary orbital energy with an efficiency described 
by parameter $\alpha_{CE}$. Formulas for both energies are given by equations A4 and
A5 of \cite{Belczynski2002}. We then take donor's mass derivative: 
\begin{equation}
\frac{dE_{\rm{bind}}}{dM_B} =-\alpha_{CE} \left( \frac{E_{\rm{orb}}}{dM_B} \right)
\label{eq.101}
\end{equation}
to obtain the second equation containing $\frac{dM_A}{dM_B}$ and $\frac{dA}{dM_B}$ 
(Eq. A7 of \cite{Belczynski2002}).
Therefore, we can rearrange the two above equations to have two ordinary differential 
equations, one for increasing mass of BH, and one for decreasing orbital separation. 
We solve them within realistic limits: using donor's envelope mass (CE) which is known 
(in contrast to integrating over unknown timescale of CE). We integrate from pre-CE 
donor mass ($M_{\rm B}$) to its post-CE mass ($M_{B,core}$) to obtain the final binary 
separation and final mass of the accreting BH. 

We assume that accretion onto a BH is always set by the Bondi rate (as implemented above). 
However, we take into account the fact that not entire infalling/accreting mass is 
actually accumulated onto a BH. Some of the accreting mass is lost before reaching BH 
(e.g., angular momentum barrier in assymetric flow around BH~\citep{Macleod2015}, 
accretion disk winds; see ~\cite{Mondal2020} and references within). We allow only some 
fraction of accreting mass to accumulate into a BH increasing its mass and spin. In particular, we estimate accretion mass $\Delta M_{\rm bondi}$ assuming that accretion proceeds with Bondi rate (i.e., integrating Eq. A9 of Belczynski et al. 2002), and we adopt that only $5\%$ of this mass actually accumulates on the BH ($\Delta M_{\rm accu}$):  

\begin{equation}
\Delta M_{\rm {accu}} = 0.05 \Delta M_{\rm {bondi}} .
\end{equation}

\end{document}